
  \magnification=1200
  \baselineskip 0.6truecm
  \def\ct {\centerline}
  
  \def\ep {\epsilon}
  \def\al {\alpha}
  \def\be {\beta}
  \def\pa {\partial}
  \def\bi {\bigskip}
  \def\de {\delta}
  \def\aa {\'a}
  \def\oo {\'o}
  \def\ee {\'e}
  \vskip 3truecm

  \ct{\bf A RELATION BETWEEN GRAVITY IN $(3+1)$--DIMENSIONS}
  \ct{\bf AND }
  \ct{\bf PONTRJAGIN TOPOLOGICAL INVARIANT}
  \vskip 1truecm
  \centerline{by}
  \vskip 1 cm
  \centerline{M. Medina and J.A. Nieto \footnote {*} {\sevenrm Research
  Supported in part by Coordinaci\oo n de Investigaci\oo n Cient\'{\i}fica de
la
  UMSNH}\footnote {}{\sevenrm Electronic address: nie@zeus.ccu.umich.mx}}
  \vskip 1 cm
  \centerline{Escuela de Ciencias F\'{\i}sico-Matem\aa ticas de la}
  \centerline{Universidad Michoacana de San Nicol\aa s de Hidalgo   }
  \centerline{Apartado postal 749, CP 58000,}
  \centerline{Morelia, Michoac\aa n,M\ee xico.}

  \bi
  \bi
  \centerline{\bf ABSTRACT}
  \vskip 1cm
   A relation between the MacDowell-Mansouri theory of gravity and the
   Pontrjagin toplogical invariant in $(3+1)$ dimensions is discussed.
   This relation may be of especial interest in the quest of finding
   a mechanism to go from non-dynamical to dynamical gravity.
   \vglue  5 cm
   August 1995
   \vfil\eject
 \baselineskip 0.6truecm{
  Topology has been a fascinating subject in mathematics. The surprise  for
  many mathematicians is that in the last few years topology has become also a
  very fascinating
  subject in theoretical physics. The main motivation to become theoretical
  physicist interested in topological aspects arises mainly from a number
  of articles written by
  Witten [1]-[4]. In particular the search for higher symmetries in the
  string theory [5]-[9] has lead Witten [10] into deeper topological
  aspects. One of the central ideas in Witten's work is to find a topological
  action from which the string theory may be derived. This means,
  roughy speaking, that starting from of a theory (topological field theory)
  with non-dynamical metric a theory (string theory) with dynamical
  metric may be derived. In spite of many attempts, however, it seems that
  this idea
  has not been completely achieved [11]-[13].
  \par
  Motivated by this idea we became intrigued that in the MacDowell-Mansouri
  theory of gravity [14]-[15] the relation between dynamical and non-dynamical
  gravity
  depends on the choice of the metric associated to the de Sitter
  (or anti-de Sitter) group. In fact, if such a metric corresponds to the
  Killing metric the action of the MacDowell-Mansouri formalism gives the
  Pontrjagin topological invariant, while if the metric is chosen to be
  proportional to the Levi-Civita tensor in four dimensions the action becomes
  Einstein-Hilbert gravitational action with cosmological constant and
  Gauss-Bonnet term included. So, a natural question is whether it is
  possible
  to find a mechanism to go from the Killing metric to the metric
  which is proportional
  to the Levi-Civita tensor in the MacDowell-Mansouri approach. If the
  answer to this question is affirmative we could in principle apply  similar
  procedure to the case of the string theory.\par
  In this article we analize a algebraic transformation which translates
  precisely the Killing metric of the  anti de Sitter group to the metric
  which is proportional to
  the Levi-Civita tensor. Such a transformation, in fact, allows a relation
  between the MacDowell-Mansouri action for gravity and the Pontrjagin
  topological invariant. \par

  Let us first briefly review the MacDowell-Mansouri theory of gravity.
  Consider the Einstein-Hilbert action  written in the tetrad formalism:
  $$
     S_1=-\int d^4\xi\ \ep^{\mu\nu\al\be}e_\mu^ae_\nu^bR_{\al\be}^{cd}
     \ep_{abcd},\eqno(1)
  $$
  where $\xi^\al$ are ``spacetime'' coordinates; $\ep^{\mu\nu\al\be}$ and
  $\ep_{abcd}$ are  Levi-Civita tensors;
  the tetrad $e_\mu^a$ is related to the metric $g_{\mu\nu}$ by the expression
  $g_{\mu\nu}=e_\mu^ae_\nu^b\eta_{ab}$ and
  $$
     R_{\mu\nu}^{ab}=\pa_ \mu w_\nu^{ab}-\pa_\nu w_\mu^{ab}
     +{1\over2}C_{efgh}^{ab} w^{ef}_\mu w^{gh}_\nu
     \eqno(2)
  $$
  is the Riemann curvature tensor written in terms of the gauge connection
  $w_\mu^{ab}$. Here $C_{efgh}^{ab}$ are the structure constants of the
  Lorentz group $SO(1,3)$. Notice, that we are using appropiate unites in
  order
  to  avoid  writting in (1) the usual constant
  factor $1\over{8\pi G}$, where $G$
  is the Newton gravitational constant.\par
  It is well know that it is possible to add to $S_1$ the cosmological
  constant term:
  $$
    S_2=\int d^4\xi\ \ep^{\mu\nu\al\be}e^a_\mu e^b_\nu e^c_\al
e^d_\be\ep_{abcd}
   . \eqno (3)
  $$
  The cosmological constant factor in (3) may arise by writting the
  tetrad $e_\mu^a$ as $\lambda{e_\mu^a}$ and by rescaling the total
  action $ S_1+S_2 $ as ${\lambda^{-2}}(S_1+S_2)$ (see ref. [15]).\par
  The central idea behind the MacDowell-Mansouri formalism is to add to
  the action $ S_1+S_2 $ the Gauss-Bonnet topological invariant
  $$
    S_0={1\over 4}\int d^4\xi\ \ep^{\mu\nu\al\be}R^{ab}_{\mu\nu}
    R^{cd}_{\al\be}\ep_{abcd}.
    \eqno (4)
  $$

  Properly combining the integrals $S_0,S_1$ and $S_2$ it is not difficult
  to show that the action $S=S_0+S_1+S_2$ may be written as
  $$
   S={1\over 4}\int d^4\xi\
\ep^{\mu\nu\al\be}\Re^{ab}_{\mu\nu}\Re^{cd}_{\al\be}
     \ep_{abcd},
    \eqno (5)
  $$
  where
  $$
     \Re^{ab}_{\mu\nu}=R^{ab}_{\mu\nu}-(e^a_\mu e^b_\nu -e^a_\nu e^b_\mu).
     \eqno (6)
  $$

  Let us now make the identification
  $$
     W^{4a}_\mu=e^a_\mu ,\ \ W^{ab}_\mu=w^{ab}_\mu .\eqno (7)
  $$
  Using these relations the curvature $\Re^{ab}_{\mu\nu}$ may be written as
  $$
     \Re^{ab}_{\mu\nu}=\pa_\mu W^{ab}_\nu -\pa_\nu W^{ab}_\mu
     +{1\over 2}C^{ab}_{{\scriptstyle EFGH}}W^{{\scriptstyle EF}}_\mu
     W^{{\scriptstyle GH}}_\nu, \eqno (8)
  $$
  where the indices {\it E,F,}$\ldots$ etc run from 0 to 4, and the only
  nonvanishing structure constants $C^{ab}_{{\scriptstyle EFGH}}$ are
  $$
   C^{ab}_{{\scriptstyle EFGH}}=\cases {C^{ab}_{efgh}&-Lorentz structure
constants,\cr
    \noalign{\vskip 1truecm}
    C^{ab}_{4f4h}= & --${1\over 2}(\de^a_f\de^b_h-\de^a_h\de^b_f).$\cr}
    \eqno (9)
  $$
  This extension of the structure constants $(C^{ab}_{efgh}\longrightarrow
  C^{ab}_{ {\scriptstyle EFGH}})$ suggests an extension of the curvature
  $$
   \Re^{ab}_{\mu\nu}\longrightarrow
   \Re^{{\scriptstyle AB}}_{\mu\nu}=\pa_\mu W^{{\scriptstyle AB}}_\nu -\pa_\nu
    W^{{\scriptstyle AB}}_\mu
    +{1\over 2}C^{{\scriptstyle AB}}_{{\scriptstyle EFGH}}W^{{\scriptstyle
EF}}_\mu
     W^{{\scriptstyle GH}}_\nu,
   \eqno(10)
  $$
  with
  $$
   \Re^{4a}_{\mu\nu}=\pa_\mu W^{4a}_\nu -\pa_\nu W^{4a}_\mu
    +C^{4a}_{4fgh}W^{4f}_\mu W^{gh}_\nu+C^{4a}_{ef4h}W^{ef}_\mu W^{4h}_\nu,
   \eqno (11)
  $$
  which in virtue of the relation $W^{4a}_\mu=e^a_\mu$ may be identified as
  the torsion
  $$
    T^a_{\mu\nu}\equiv\Re^{4a}_{\mu\nu}=\pa_\mu e^a_\nu-\pa_\nu e^a_\mu
    +w^a_{\mu b}e^b_\nu-w^a_{\nu b}e^b_\mu .
    \eqno (12)
  $$
  The quantities
  $$C^{{\scriptstyle AB}}_{{\scriptstyle EFGH}}={1\over 2}
    [\de^{{\scriptstyle A}}_{{\scriptstyle E}}\de^{{\scriptstyle
B}}_{{\scriptstyle H}}
    \eta_{{\scriptstyle FG}}-
    \de^{{\scriptstyle A}}_{{\scriptstyle E}}\de^{{\scriptstyle
B}}_{{\scriptstyle G}}
    \eta_{{\scriptstyle FH}}-
    \de^{{\scriptstyle A}}_{{\scriptstyle F}}\de^{{\scriptstyle
B}}_{{\scriptstyle H}}
    \eta_{{\scriptstyle EG}}+
   \de^{{\scriptstyle A}}_{{\scriptstyle F}}\de^{{\scriptstyle
B}}_{{\scriptstyle G}}
    \eta_{{\scriptstyle EH}}]-{1\over 2}
   [{{\scriptstyle A}}\leftrightarrow {{\scriptstyle B}}],
   \eqno (13)
  $$
  may be identified, now, with the structure constants of the
  anti-de Sitter group $S(2,3)$ (a similar
  result may be obtained in the case of the de Sitter group $SO(1,4)$).\par
  Finally, let us introduce the quantity
  $$
    g_{{\scriptstyle ABCD}}=\cases{g_{abcd}=\ep_{abcd},\cr
    \noalign{\vskip 1 truecm}g_{4a4b}=-\eta_{ab},\cr \noalign{\vskip 1 truecm}
    0\ otherwise. \cr }\eqno (14)
  $$
  Notice that $g_{{\scriptstyle ABCD}}=-g_{{\scriptstyle
BACD}}=-g_{{\scriptstyle ABDC}}
  =+g_{{\scriptstyle CDAB}}$.\par
  Making the torsion $T^a_{\mu\nu}=0$ we find that the action (4) may
  be written as
  $$
    S={1\over 4}\int d^4\xi\ \ep^{\mu\nu\al\be}\Re^{{\scriptstyle AB}}_{\mu\nu}
     \Re^{{\scriptstyle CD}}_{\al\be}g_{{\scriptstyle ABCD}}.  \eqno (15)
  $$
  An important feature of this action is that it is independent of the metric
  $g_{\mu\nu}$ and the Christoffel simbols $\Gamma^\mu_{\al\be}$ and it
  depends only on the gauge field $W^{{\scriptstyle AB}}_\mu=-W^{{\scriptstyle
BA}}_\mu$
  associated to the
  anti-de Sitter group $SO(2,3)$ (or the de Sitter group $SO(1,4))$.\par
  The action (15) is so similar to the Pontrjagin topological invariant
  action
  $$
    {\cal S}={1\over 8}\int d^4\xi\ \ep^{\mu\nu\al\be}{\cal R}^{{\scriptstyle
AB}}_{\mu\nu}
    {\cal R}^{{\scriptstyle CD}}_{\al\be}
     G_{{\scriptstyle ABCD}},
    \eqno (16)
  $$
  where
  $$
   {\cal R}^{{\scriptstyle AB}}_{\mu\nu}=\pa_\mu{\cal W}^{{\scriptstyle
AB}}_\nu-
   \pa_\nu{\cal W}^{{\scriptstyle AB}}_\mu
   +{1\over 2}f^{{\scriptstyle AB}}_{{\scriptstyle EFGH}}{\cal
W}^{{\scriptstyle EF}}_\mu
   {\cal W}^{{\scriptstyle GH}}_\nu
   \eqno (17)
  $$
  and
  $$
    G_{{\scriptstyle ABCD}}=(\eta_{{\scriptstyle AC}}\eta_{{\scriptstyle BD}}-
    \eta_{{\scriptstyle AD}}\eta_{{\scriptstyle BC}})
    \eqno (18)
  $$
  that we become intrigued if there is some kind of transformation
  $$
  \Re^{{\scriptstyle AB}}_{\mu\nu}\leftrightarrow{\cal R}^{{\scriptstyle
AB}}_{\mu\nu}
  \eqno (19)
  $$
  and
  $$
    g_{{\scriptstyle ABCD}}\leftrightarrow G_{{\scriptstyle ABCD}}\eqno(20)
  $$
  such that
  $$
   S\leftrightarrow {\cal S} .\eqno(21)
  $$
  In the expression (17) the quantities $f^{{\scriptstyle AB}}_{{\scriptstyle
EFGH}}$
  are the structure
  constants related, of course, to the
  anti-de Sitter group (or the de Sitter group).\par
  In order to find such a transformation we first expand (15) and (16)
  as follows:
  $$
    S={1\over 4}\int d^4\xi\ \ep^{\mu\nu\al\be}\Re^{ab}_{\mu\nu}
   \Re^{cd}_{\al\be}\ep_{abcd}
       -\int d^4\xi\ \ep^{\mu\nu\al\be}T^{a}_{\mu\nu}T^{b}_{\al\be}\eta_{ab}
    \eqno (22)
  $$
  and
  $$
    {\cal S}={1\over 8}\int d^4\xi\
\ep^{\mu\nu\al\be}{\cal R}^{ab}_{\mu\nu}{\cal R}^{cd}_{\al\be}
        (\eta_{ac}\eta_{bd}-\eta_{ad}\eta_{bc})-
 {1\over 2}\int d^4\xi\
\ep^{\mu\nu\al\be}{\cal R}^{4a}_{\mu\nu}{\cal R}^{4b}_{\al\be}\eta_{ab}.
    \eqno (23)
  $$
  Of course, in the MacDowell-Mansouri approach $T^{a}_{\mu\nu}=0$,
  but for the moment let us consider that $T^{a}_{\mu\nu}\ne 0$. At this
respect
  we should mention important aspects. Since we are considering here
  pure gravity the necessity to impose the constraint $T^{a}_{\mu\nu}=0$
  arises  if we want the action (22) to be consistent with Einstein`s
  gravitational therory. A remarkable feature of the action (22) is that
  by eliminating the second integral, the first integral precisely reproduce
  the constraint $T^{a}_{\mu\nu}=0$ under variation. We note, however,
  that this compatibility holds at the classical level. This is an
  important observation which should be carefully considered at the
  quantum level (see ref. [14] for more details).\par
  From the
  two expressions (22) and (23) we see that in principle $T^{a}_{\mu\nu}$
  may be identified
  with ${\cal R}^{4a}$ and the first term in $S$ may be identified with the
  first term in ${\cal S}$. So, let us first concentrate our attention in
  the integrals:
  $$
    \hat S={1\over 4}\int d^4\xi\
\ep^{\mu\nu\al\be}\Re^{ab}_{\mu\nu}\Re^{cd}_{\al\be}
   \ep_{abcd}\eqno (24)
  $$
  and
  $$
    \hat{\cal S}={1\over 8}\int d^4\xi\ \ep^{\mu\nu\al\be}{\cal
R}^{ab}_{\mu\nu}
    {\cal R}^{cd}_{\al\be}(\eta_{ac}\eta_{bd}-\eta_{ad}\eta_{bc}).
    \eqno (25)
  $$
  In order to find a relation between these two integrals it is convenient
  to introduce the quantity
  $$
    {}^\pm N^{ab}_{cd}={1 \over 2}(\delta^{ab}_{cd}\pm \ep^{ab}_{\ \ cd}),
    \eqno(26)
  $$
  where
  $$
    \delta^{ab}_{cd}=\delta^{a}_{c}\delta^{b}_{d}-\delta^{a}_{d}\delta^{b}_{c}.
  \eqno(27)
  $$
  We can check that this quantity satisfies the relations
  $$
    {1 \over 2}{}^\pm N^{ab}_{ef}{}^\pm N^{cd}_{gh}
    (\eta_{ac}\eta_{bd}-\eta_{ad}\eta_{bc})=\pm \ep_{efgh},
    \eqno (28)
  $$
  $$
    {}^+N^{ab}_{cd}{}^-N^{cd}_{ef}=\delta^{ab}_{ef}.
    \eqno (29)
  $$
  Thus, using (28) we find that $\hat S$ becomes
  $$
    \hat S={1 \over 8}\int d^4\xi\ \ep^{\mu\nu\al\be}\Re^{ab}_{\mu\nu}
   \Re^{cd}_{\al\be}
     {}^+N^{ef}_{ab}{}^+N^{gh}_{cd}(\eta_{eg}\eta_{fh}-\eta_{eh}\eta_{fg}).
    \eqno (30)
  $$
  Terefore, the problem to obtain $\hat{\cal S}$ from $\hat S$ may be
  accomplished if we
  consider the transformation
  $$
    {\cal R}^{ab}_{\mu\nu}={}^+N^{ab}_{cd}\Re^{cd}_{\mu\nu},
    \eqno (31)
  $$
  because then
  $$
  \hat S =\hat {\cal S}.\eqno(32)
  $$
  However, the solution is not so simple because in order to have (31) we
  need to find the relations;
  $$w^ {ab}_\mu \mapsto  {\cal W}^{ab}_\mu ,$$
  $$\eqno(33)$$
  $$w^ {4a}_\mu \mapsto  {\cal W}^{4a}_\mu ,$$
  and
  $$C^{ab}_{{\scriptstyle EFGH}} \mapsto  f^{ab}_{{\scriptstyle
EFGH}}.\eqno(34)$$
  From the definitions of ${\cal R}^{ab}_{\mu\nu}$ and $R^{ab}_{\mu\nu}$ we
find
  that (31) follows if the relations
  $$
    {\cal W}^ {ab}_\mu={}^+N^{ab}_{cd}w^ {cd}_\mu ,\eqno (35)
  $$
  $$
    f^{ab}_{efgh}={1\over
4}{}^+N^{ab}_{cd}C^{cd}_{mnij}{}^-N^{mn}_{ef}{}^-N^{ij}_{gh}
   , \eqno (36)
  $$
  $$
   f^{ab}_{4f4h}={1\over 2}{}^+N^{ab}_{cd}C^{cd}_{4f4h},\eqno (37)
  $$
  and
  $$
   {\cal W}^{4a}_\mu=\sqrt 2 \ w^{4a}_\mu \eqno (38)
  $$
  are satisfied. The relation (35) follows directly from (31). The
  expressions (36), (37), and (38) , however, are more difficult to
  obtain and require special attention to numerical factors. We note
  that in order to obtain (36) we used (29). It is interesting to observe
  the important role  played by the quantity (26) and the relations
  (28) and (29); without these relations to find the expressions
  (35)-(36) would be very much difficult.

  It remains to clarify the meaning of the structure constants
  $f^{{\scriptstyle AB}}_{{\scriptstyle EFGH}}$.
  What we know is that $C^{{\scriptstyle AB}}_{{\scriptstyle CDEF}}$ are the
structure
  constant of the   anti-de Sitter group $SO(3,2)$ whose generators
  $S_{{\scriptstyle AB}}$ satisfy the  algebra:
  $$
   [S_{{\scriptstyle EF}},S_{{\scriptstyle GH}}]=C^{{\scriptstyle
AB}}_{{\scriptstyle EFGH}}
   S_{{\scriptstyle AB}}.\eqno (39)
  $$
  This algebra can be broken as follows
  $$
   [S_{ef},S_{gh}]=C^{ab}_{efgh}S_{ab},\eqno (40)
  $$
  $$
   [S_{4f},S_{gh}]=2C^{4b}_{4fgh}S_{4b},\eqno (41)
  $$
  $$
   [S_{4f},S_{4h}]=C^{ab}_{4f4h}S_{ab}. \eqno (42)
  $$
  The first bracket (40) may be multiplied by
  ${1\over 4}{}^-N^{ef}_{ij}{}^-N^{gh}_{kl}$
  in order to obtain
  $$
    [{}^-S_{ij},{}^-S_{kl}]={1\over 4}{}^+N^{ab}_{rs}C^{rs}_{efgh}
    {}^-N^{ef}_{ij}{}^-N^{gh}_{kl}{}^-S_{ab},
    \eqno (43)
  $$
  where we used (29) and the definition
  ${}^-S_{ij}={1\over 2}{}^-N^{kl}_{ij}S_{kl}$.
  From this expression we see that the structure constants $f^{ab}_{efgh}$
  given in (36) precisely corresponds to the factor in front of ${}^-S_{ab}$
  in (43). In fact, using (36) we get
  $$
   [{}^-S_{ij},{}^-S_{kl}]=f^{rs}_{ijkl}{}^-S_{rs}. \eqno (44)
  $$
  Similarly, multiplying (41) by ${1\over 2\sqrt 2}{}^-N^{gh}_{ij}$
  and defining ${}^-S_{4a}={1\over\sqrt 2}S_{4a}$ we find
  $$
    [{}^-S_{4f},{}^-S_{ij}]=2({1\over 2}{}^-N^{gh}_{ij}C^{4b}_{4fgh}{}^-S_{4b})
     \eqno (45)
  $$
  which suggests to write
  $$
    f^{4b}_{4fij}={1\over 2}{}^-N^{gh}_{ij}C^{4b}_{4fgh}.  \eqno (46)
  $$
  Further, the bracket (42) can be written as
  $$
   [{}^-S_{4f},{}^-S_{4h}]={1\over 2}C^{ab}_{4f4h}{}^+N^{ij}_{ab}{}^-S_{ij},
   \eqno (47)
  $$
  with
  $$
    f^{ij}_{4f4h} ={1\over 2}{}^+N^{ij}_{ab}C^{ab}_{4f4h}.\eqno (48)
  $$

  Therefore, the anti-de Sitter algebra (39) may be rewritten as
  $$
    [{}^-S_{{\scriptstyle AB}},{}^-S_{{\scriptstyle CD}}]=
    f^{{\scriptstyle EF}}_{{\scriptstyle ABCD}}{}^-S_{{\scriptstyle EF}}, \eqno
(49)
  $$
  where ${}^-S_{ab}={1\over 2}{}^-N^{ef}_{ab}S_{ef}$ and
  ${}^-S_{4a}={1\over\sqrt 2}S_{4a}$.
  Consequently (49) implies that the structure constants $f^{{\scriptstyle
EF}}_{{\scriptstyle ABCD}}$
  also correspond to the anti-de Sitter group.\par
  With all these results we can see that, the connection between the integrals
  $$
    \hat S_T=-\int d^4\xi\ \ep^{\mu\nu\al\be}T^a_{\mu\nu}T^b_{\al\be}\eta_{ab}
    \eqno (50)
  $$
  and
  $$
    \hat {\cal S}_T=-{1\over 2}\int d^4\xi\ \ep^{\mu\nu\al\be}{\cal
R}^{4a}_{\mu\nu}
    {\cal R}^{4b}_{\al\be}\eta_{ab} \eqno (51)
  $$
  is straightforward. In fact, we find
  $T^a_{\mu\nu}={1\over\sqrt 2}{\cal R}^{4a}_{\mu\nu}$.\par
  To conclude let us make the following comments. The message given by this
  work may be expressed as follows: Start with Einstein-Hilbert action, add a
  cosmological constant term and  Gauss-Bonnet
  topological invariant form, transform each
  term according to (31) and finally add a torsion term breaking reflexion
  what we get it is the Pontrjagin topological invariant. Or inversely,
  starting with Pontrjagin topological invariant make the torsion vanishes
  then make the transformation (31) and eliminate the Gauss-Bonnet
  topological invariant and the cosmological
  constant what we get it is the Eistein-Hilbert action. In other words, the
  transition from
  a non-dynamical to dynamical gravity depends of making the torsion zero and
  making appropriate transformation using ${}^+N^{ab}_{cd}$. Finally,
  it is interesting that all this mechanism has been derived
  classically. Of course, It will be very interesting to find a similar
  dynamical mechanism and to exploite such a dynamical mechanism
  at the quantum level. But at the present time we do not see how to
  achieve this goal. Nevertheless, we think that the present work may
  be very useful in that direction.
  Further, since the Pontrjagin topological
  invariant may be related to Chern-Simons of gauge group
  $SO(2,3)$ which at
  the same time is related to $(1+1)$-conformal field theory [16]-[19]
  we think our work
  may be thought as bridge between gravity in $(3+1)$-dimensions and
  $(1+1)$-conformal
  field theory. Since $(1+1)$-conformal theory is closely related to string
  theory it seems that our work suggests that string theory may be obtained
  from gravity!\par
  \vfill\eject
  {\bf References}
  \bi
  \bi
  \settabs 4 \columns
  \item{[1]} E. Witten, Comm. Math. Phys. 117 (1988) 353.
  \item{[2]} E. Witten, Comm. Math. Phys. 121 (1989) 351.
  \item{[3]} E. Witten, Nucl.Phys. B311 (1988) 46.
  \item{[4]} E. Witten, Nucl. Phys. B323 (1989)  113.\par
  \item {[5]} M. B. Green, J. H. Schwarz, and E. Witten, Superstring Theory,
     Vol I and II  (Cambridge, 1987).
  \item{[6]} D. Lust and S. Theisen, Lectures on String Theory
   (Springer-Verlag, 1989).
  \item{[7]} M. Kaku, Introduction to Superstrings (Springer-Verlag, 1988);
   Strings, Conformal Fields, and Topology (Springer-Verlag, 1991).
  \item{[8]} B. Hatfield, Quantum Field Theory of Point Particles and
    Strings (Addison-Wesley, 1992).
  \item{[9]} J. Polchinski, Joe's Big Book of String (Cambridge, 1995).
  \item{[10]} E. Witten, ``The Search for Higher Symmetry in String Theory'',
   Lecture given at Mtg. on String Theory of the Royal
   Society, London, England, Dec 1988. In London 1988, Proceedings, Physics
   and mathematics of strings 31-39
   and Inst. Adv. Stud. Princeton-IASSNS-HEP-88-55
   ( Philos. Trans. R. Soc. London A320 (1989) 349-357).\par
  \item {[11]} E. Witten, ``Two-Dimensional String Theory and Black Holes'',
   Lectures given at Conf. on Topics
  in Quantum Gravity, Cincinnati, OH, Apr. 3-4 1992; In Mansouri, F. (ed),
  Scanio, J. J. (ed.): Quantum gravity and beyond, hep-th/9206069;
  "On Black Holes in String Theory", Lecture given at Strings'91 Conf.
  Stony Brook, N. Y., Jun. 1991. Published in Strings: Stony Brook 1991:
  184-192. \par
  \item{[12]} E. Witten, Phys. Rev. D47, 3405 (1993).\par
  \item{[13]} E. Witten, Published in Salamfest 257 (1993). hep-th/9306122.\par
  \item {[14]} S. W. MacDowell and F. Mansouri, Phys. Rev. Lett 38 (1977) 739;
  Mansouri Phys. Rev. D16 (1977) 2456.
  \item{[15]} P.G.O. Freund, Introduction to Supersymmetry (Cambridge, 1988).}
  \item{[16]} A. A. Belavin , A. M. Polyakov, and A. B. Zamolodchikov,
  Nucl. Phys. B241 (1984) 333.
  \item{[17]} D. Friedan , E. Martinec, and S. Shenker, Nucl. Phys. B271
   (1986) 93.
  \item{[18]} P. Ginsparg, in Fields, Strings, and Critical Phenomena,
   Les Houches Session XLIX, edited by  E. Brezin and J. Zinn-Justin
   (Elsevier, 1989) 1.
  \item{[19]} J. L. Cardy, in Fields, Strings, and Critical Phenomena,
   Les Houches Session XLIX, edited by E. Brezin and  J. Zinn-Justin
   (Elsevier, 1989) 169.
  \bye